\documentstyle{article}
\begin{document}

%This is the cover page
\null
\vspace{2cm}
\begin{center}
\begin{Large}
\begin{bf}
Determining Gravitational Masses of Galaxy Clusters With\\
(1)Virial Equilibrium And (2)Arc-like Images\\
\end{bf}
\bigskip
\bigskip
\bigskip
\bigskip

Xiang-Ping Wu$^{1,2}$ \\
\end{Large}
\begin{large}

\bigskip
\bigskip

$^{1}$Beijing Astronomical Observatory,\\
Chinese Academy of Sciences, Beijing 100080, China\\
%\bigskip
%and\\
\bigskip
%$^{2}$DAEC, Observatoire de Paris-Meudon, 92195 Meudon Principal Cedex,
%         France\\
and\\
\bigskip
$^{2}$Department of Physics, Tianjin Normal University, Tianjin 300074, China\\

\end{large}

\bigskip
\bigskip
\bigskip
\bigskip
\bigskip
\bigskip

\begin{large}
Submitted to\\

\bigskip
\bigskip

{\bf The Astrophysical Journal (Letters)}\\

\bigskip
\bigskip
\bigskip

\end{large}

\end{center}

\bigskip
\bigskip
\bigskip
{\noindent}-------------------------------------------------------------------------------------\\
{\noindent}Present address: Xiang-Ping Wu$^{1}$\\
{\noindent}E-mail: wxp@mesiob.obspm.fr ~ or ~ wxp@bao01.bao.ac.cn

\newpage

\begin{center}
\begin{Large}
\begin{bf}
Determining Gravitational Masses of Galaxy Clusters  With\\
(1)Virial Equilibrium And (2)Arc-like Images \\
\end{bf}
\end{Large}

\bigskip
\bigskip

\begin{Large}
Xiang-Ping Wu \\

\bigskip
\begin{large}
Beijing Astronomical Observatory, Chinese Academy of Sciences,\\
Beijing 100080, People's Republic of  China\\
and\\
Department of Physics, Tianjin Normal University, Tianjin 300074, China\\
\end{large}
\bigskip
\bigskip
\bigskip
\bigskip
\bigskip
\bigskip
\bigskip

{\bf ABSTRACT}\\
\end{Large}
\end{center}

The mass derived from gravitational lensing reflects
the total mass contained in the lensing system, independent of the
specific matter contents and states.
A comparison of
the dynamical masses from hydrostatic equilibrium with
the gravitational masses from arc-like images of background galaxies
is made for four clusters of galaxies at intermediate
redshits. It is found that virial analysis has underestimated
the total cluster masses (from lensing) by a factor of $3\sim6$ within
a radius of $\sim0.3$ Mpc $h_{50}^{-1}$ around the cluster centers,
indicating that clusters of galaxies
might not be regarded as the well relaxed virialized systems.
The increase of the total cluster masses obtained from lensing
leads to the decrease of the baryon fractions of clusters of galaxies,
which  provides a crue for solving the ``$\Omega_0$ disprepancy
puzzle"  in cosmology.\\

\bigskip
\bigskip
\bigskip
\bigskip
{\noindent}-----------------------------------------------------------------------\\
{\noindent}{\bf\it Subject headings:} ~~galaxies: clustering --
gravitational lensing -- dark matter -- cosmology\\

%--------------------------------------------------------
\newpage
\begin{large}

\begin{center}
\begin{Large}
{\bf 1. ~INTRODUCTION}\\
\end{Large}
\end{center}

Determination of total masses gravitationally bounded in clusters
of galaxies is important for our understanding of the nature and
evolution of structures on large scales. Traditionally, the gravitational
masses of clusters of galaxies are estimated from the dynamical analysis
assuming an hydrostatic equilibrium of both the hot intracluster gas and
the galaxies with the binding cluster potential (Cavaliere \&
Fusco-Femiano 1976, Jones \& Forman 1984;
Cowie, Henriksen \& Mushotzky 1987; Hughes 1989). Particularly,
sufficient data are now available in X-ray (e.g., the EINSTEIN Observatory,
EXOSAT, ROSAT, etc.) so that the virial masses of some nearby galaxy clusters
(e.g., Coma, A 2256, Perseus) can be computed very precisely
using the temperature and the density of the
hot X-ray emitting diffuse gas (Eyles et al. 1991; Arnaud et al. 1992;
Briel, Henry \& Bohringer 1992; Miyaji et al. 1993; Henry, Briel \& Nulsen
1993).  This straightforward method for computing the total dynamical
mass from the X-ray emitting gas reveals a large amount of unseen matter
concentrated in the central regions of clusters. \\

The difficulty of the standard hydrostatic cluster model, however, comes from
the so-called ``$\Omega_0$ discrepancy problem". The fraction of gas
to total mass, i.e., the baryonic fraction,
($\Omega_b/\Omega$) obtained from the virial theorem in clusters of
galaxies is much higher than one would obtain from the primordial
nucleosynthesis (Briel, Henry \& Bohringer 1992; While et al. 1993),
if the matter in clusters is representive of the Universe. The similar
discrepancy has been also found in small groups of galaxies (Ponman \&
Bertram 1993;  Mulchaey et al. 1993; Henriksen \& Mamon 1994).
Adopting the new measurement of $\Omega_b=0.022 h_{50}^{-2}$ from the
abundance of $D/H$ in primordial gas
(Songaila et al. 1994; Carswell et al. 1994)
and taking typical value of $\Omega_b/\Omega_0\sim0.15 h_{50}^{-3/2}$
within $3$Mpc $h_{50}^{-1}$ of clusters of galaxies from the virial
equilibrium (Ceballos \& Barcons 1994), one would simply expect
a very low density Universe of $\Omega_0\approx0.15$, while the inflation model
and the large-scale motions of galaxies have led to $\Omega_0\approx1$.
This discrepancy has been an unsolved puzzle in today's astrophysics although
some possibilities have been suggested (White et al. 1993).\\

The crucial point in the above estimate of baryon fractions of clusters of
galaxies is the hypothesis that the clusters of galaxies were thought to be
in the state of dynamical equilibrium. On the other hand, the requirement of
the cosmic nucleosynthesis and the argument of $\Omega_0\approx1$
imply that the virial analysis has probably underestimated the total masses of
clusters of galaxies. One of the ways to test this question is to have
another independent method of determining the total gravitational masses of
clusters of galaxies. \\

The arc-like images of faint background galaxies associated with the
gravitational potential of foreground galaxy clusters open a new
way to find the binding gravitational masses of galaxy clusters,
independent of the state equations of
the hot X-ray gas, the galaxies and the unseen matter.
Modelling  giant arcs and arclets has provided a direct
measurement of the projected masses interior to the arcs
(Soucail \& Mellier 1993; references therein),
while using the properties of giant luminous arcs (position, length,
width, axial ratio, number, etc.) has indicated the compact distributions of
dark matter in the cores of arc clusters (Hammer 1991; Wu \& Hammer 1993;
 etc.).  In particular, a comparison of the masses derived
from giant luminous arcs with the masses from the viral theorem has been
made for a few clusters by Fort (1992). He found that the total masses
derived from the two independent methods are essentially consistent.
However, it is noticed that Fort's comparisons were made within
the cores of clusters,  which may suffer from the contaminations of
cooling flow.\\

This letter presents the results of the total masses for four clusters
of galaxies at intermediate redshifts
utilizing two independent methods by solving
(1)the equation of hydrostatic equilibrium for the X-ray gas  and
(2)the equation of gravitational lensing  for the observed arcs associated
with the clusters.  These arcs are carefully chosen so that they are
well separated from the X-ray gas cores
($\sim0.25$ Mpc $h_{50}^{-1}$) (Abramopoulos \& Ku 1983; Jones \& Forman 1984)
and the cooling radii ($\sim0.13$ Mpc $h_{50}^{-1}$) (Edge \& Stawart 1991)
of galaxy clusters, and would then reflect the masses outside the central
regions of the clusters.  A direct test for the plausibility of the assumption
that the galaxy clusters are the well relaxed virialized systems will be
given by comparing the total masses obtained from the two different methods.\\

\bigskip
\bigskip

\begin{center}
\begin{Large}
{\bf 2. ~DYNAMICAL MASS FROM VIRIAL EQUILIBRIUM}\\
\end{Large}
\end{center}

Assuming that both the X-ray gas and the galaxies are in hydrostatic
equilibrium with the binding gravitational potential of a spherical
cluster,   one can write the virial mass of the  cluster within radius
$r$ as (Cowie, Henriksen \& Mushotzky 1987)
%1
\begin{equation}
M_v(r)=-\frac{kTr}{G\mu m_p}\left(\frac{d\ln n}{d\ln r}+
\frac{d\ln T}{d\ln r}\right),
\end{equation}
where $n$ and $T$ are the gas density and temperature, respectively, $\mu$
is the mean particle weight in units of the proton mass $m_p$. The spatial
distributions of $n$ and $T$ in cluster are obtained by inverting
the observed X-ray surface brightness profile which is usually well-fitted
by the $\beta$ model
%2
\begin{equation}
S(\theta)=S_0\left[1+(\theta/\theta_c)^2\right]^{1/2-3\beta}
\end{equation}
with a core radius of $\theta_c$. Note that $M_v$ is independent of
the central gas density or the central surface brightness $S_0$. In the
case of isothermal gas distribution which has been found to be consistent
with the observations of some nearby clusters (e.g., A 2256, see
Henry, Briel \& Nulsen 1993) and which will be adopted in the following
calculations,  the viral mass [eq.(1)] and  its corresponding mass density
profile are
%3
\begin{equation}
M_v(x)=3\beta\frac{kTr_c}{G\mu m_p}\frac{x^3}{1+x^2}
\end{equation}
and
%4
\begin{equation}
\rho(x)=\frac{3\beta}{4\pi}\frac{kT}{G\mu m_p}\frac{1}{r_c^2}
\frac{3+x^2}{(1+x^2)^2},
\end{equation}
respectively,
where $r_c$ is the core radius of the X-ray gas corresponding to $\theta_c$,
and $x=r/r_c$. \\

To compare with the mass estimated from the arc-like
images using gravitational lensing in next section, one needs to find
the projected virial mass $m_v(b)$ within a radius of $b$ on
the cluster plane. A  calculation utilizing eq.(4) yields
%5
\begin{equation}
m_v(b)=1.14\times10^{14}\;\beta\;\tilde{m}(b)\;
\left(\frac{kT}{\rm keV}\right)\left(\frac{r_c}{\rm Mpc}\right)\;M_{\odot}
\end{equation}
%6
\begin{equation}
\tilde{m}(b)=\frac{R_0^3}{1+R_0^2}-\int_{b_0}^{R_0}x\sqrt{x^2-b_0^2}
\frac{3+x^2}{(1+x^2)^2}\;dx.
\end{equation}
Here $b_0=b/r_c$ and $R_0=R/r_c$. R is the physical size of the cluster.
The numerical computation shows that $m(b)$ remains almost unchanged
for $R$ ranging from 3 Mpc to 100 Mpc, $r_c\sim 0.25 Mpc$ and
$b\sim r_c$.  In the following calculations $R$ will be taken to be
$5$ Mpc $h_{50}^{-1}$. \\

In the case of where the temperature of the cluster is unknown, the strong
correlation between the X-ray luminosity $L_x$ and the temperature $T$
found by Edge \& Steward (1991) is used
%7
\begin{equation}
(T/{\rm keV})=10^{-12.73}(L_x/erg\;s^{-1})^{0.30}.
\end{equation}
Furthermore, a mean core radius of $r_c=0.25$ Mpc $h_{50}^{-1}$ and
a value of $\beta=2/3$ will be adopted for the cluster X-ray surface
brightness distribution (Abramopoulos \& Ku 1983; Jones \& Forman 1984;
Henry et al. 1992).

\bigskip
\bigskip

\begin{center}
\begin{Large}
{\bf 3. ~GRAVITATIONAL MASS FROM ARC-LIKE IMAGES}\\
\end{Large}
\end{center}

Assuming  a spherical matter distribution of cluster of galaxies
in the lensing equation,
one can write the gravitational mass of a galaxy cluster projected
within a radius of $b$ or $\theta$ (in arcseconds) on the cluster plane as
%8
\begin{equation}
m_g(b)=7.37\times10^{11}\;
(\theta-\theta_0)\theta\;\frac{d_s d_d}{d_{ds}}\;M_{\odot}h_{50}^{-1}
%m_g(b)=\frac{c^2}{4G}(b-\ell)b\frac{D_s}{D_d D_{ds}},
\end{equation}
where $\theta_0$ is the alignment parameter (in arcseconds) of the lensing
cluster, $d_d$, $d_s$ and $d_{ds}$ are the angular diameter distances
in units of $(c/H_0)$ to
the lensing cluster, to the background source (galaxy) and from
the cluster to the source, respectively.  Actually, the undisturbed
position of the background source, $\theta_0$, is unmeasurable.
However, one can estimate its value through the length ($L$) and the axial
ratio ($L/W$) of the arc-like image (Wu \& Hammer 1993)
%9
\begin{equation}
\theta_0=\frac{L}{2(L/W)\sin(L/2b)}.
\end{equation}
If the redshifts of the arc-like images have not been presently measured,
a low limit can be set on the mass of their associated cluster
%10
\begin{equation}
m_{g,min}(\theta)=7.37\times10^{11}\;
(\theta-\theta_0)\theta\; d_d\;M_{\odot}h_{50}^{-1}.
\end{equation}

\bigskip
\bigskip

\begin{center}
\begin{Large}
{\bf 4. RESULTS FOR FOUR CLUSTERS OF GALAXIES}\\
\end{Large}
\end{center}

The arc-like images are chosen so that their distances from the centers
of their associated clusters are larger than the cooling radii which are
estimated to be $0.13$ Mpc $h_{50}^{-1}$ on average using the EXOSAT
X-ray data (Edge \& Steward 1991).  This selection guarantees that
the virial analysis does not suffer from the contaminations from
the possible cooling flow.  Considering the fact that the core radius of
the X-ray gas of cluster is typically $0.25$ Mpc $h_{50}^{-1}$, one further
requires
that the arc-like images locate outside the central core of the cluster.
As a result, four clusters of galaxies at intermediate redshift are chosen
in which the distortions of background galaxies have been detected:
Abell 370 (A5), MS 1006.0+1202 (arc 4), MS 1008.1-1224 (arc 2)
and MS 1910.5+6736.  The first example  A5 in Abell 370 is adopted
from Fort et al. (1988) and the rest three from the arc survey in
the distant EMSS clusters (Hammer et al. 1993; Le F\`evre et al. 1994).
The X-ray luminosity for A 370 is taken from Henry et al. (1982) and
the X-ray data for the rest three are chosen from the new EMSS catalog
(Gioia \& Luppino 1994).  Unfortunately, the redshift data are not
available for all the three arcs selected in the EMSS sample while
the  A5 in A 370 locates probably at $z_s\approx1.3$.\\

The properties of the four arc-cluster systems  and
the results of the clusters masses obtained using eq.(5) and eq.(8)
are given in Table 1. Two redshift values of $z_s=0.6$ and $z_s=2$ have been
applied for the three arcs detected in the EMSS sample, which cover
the most possible redshift range of the arc-like images of faint
background galaxies (Wu \& Hammer 1993). It appears that a significant
difference between the virial masses and the gravitational masses from
arc-like images is shown in all the four arc-cluster systems: the dynamical
method from  hydrostatic equilibrium results in the cluster masses
3 to 6 times smaller than the masses from the method of gravitational lensing
within the circle of $0.28$--$0.35$ Mpc $h_{50}^{-1}$ around the clusters
centers. This mass discrepancy is so large and common that it is very hard
to attribute the excess of gravitational masses from lensing to some
errors of observations (the X-ray data, arc-like image data, etc.) and
some scatters of the correlation of $L_x$ with $T$ [eq.(7)] (Edge \&
Stewart 1991).
It is then very likely that the virial equilibrium has underestimated
the total gravitational masses of clusters of galaxies by a factor of
at least 3 outside the X-ray gas radii. \\

\bigskip
\bigskip

\begin{center}
\begin{Large}
{\bf 5. DISCUSSION AND CONCLUSIONS}\\
\end{Large}
\end{center}

The gravitational mass derived from gravitational lensing reflects the
total mass of the lensing system, despite whether or not the system is in
the state of hydrostatic equilibrium. The cluster masses obtained from
the arc-like images of background galaxies can then be regarded as
the most reliable values. Utilizing this method for four arc-cluster
systems  leads to the conclusion that the previous dynamical
analysis from virial theorem has produced
relatively lower cluster masses than the gravitational masses from
arc-like images, indicating that clusters of galaxies
might not be taken as the well relaxed virialized systems. \\

It is noticed that the four galaxy clusters used in the computations are all
at intermediate redshifts ($z_d=0.2$--$0.4$) and have strong X-ray
emissions ($L_x>4\times10^{44}\; erg/s$).  These X-ray selected distant
galaxy clusters may have undergone a significant evolution
with cosmic epoch (Gioia et al. 1990; Edge et al. 1990; Henry et al. 1992)
although the scenario of strong X-ray luminosity evolution
gives rise to the few clusters of galaxies
as the hosts for the observed giant luminous arcs (Wu 1993).
It is then very likely that the galaxy clusters having strong
X-ray emission are still in the non-virialized cosmo-dynamical state.
This indeed contradicts with the recent claim by Lubin \& Bahcall (1993)
that both the gas and galaxies are in hydrostatic equilibrium with the
binding cluster potential,  based on the statistical
analysis of the observed galaxy velocity dispersion and the X-ray
temperature of the hot gas in clusters of galaxies. \\

A consequence of replacing the virial masses by the gravitational masses
from lensing for clusters of galaxies would
provide a new clue for resolving the ``$\Omega_0$ discrepancy puzzle".
The baryon fractions of clusters of galaxies can now be written as
%11
\begin{equation}
\frac{\Omega_b}{\Omega_0}\;=\;\frac{m_b(b)}{m_v(b)}\;\;\frac{m_v(b)}{m_g(b)},
\end{equation}
where the term $m_b(b)/m_v(b)$ is the estimate using the virial
mass while the term $m_v(b)/m_g(b)$ gives the correction factor from
gravitational lensing that reduces the previous value
$m_b(b)/m_v(b)$ by a factor of $3$--$6$ at $b\approx0.3$ Mpc $h_{50}^{-1}$.
The application of the above expression is, however,
limited to the innerparts of the locations of arc-like images in clusters,
which  prevents from an estimate of $\Omega_b/\Omega_0$ over all the
regions of clusters. It then remains unclear if the baryon fractions
of clusters of galaxies would decrease  with the increase of
radii.  If one adopts a factor of $\sim4$ for
$m_g(0.3{\rm Mpc})/m_v(0.3{\rm Mpc})$, the previous estimate of baryon
fractions of clusters of galaxies should be decreased by the same factor,
leading to $\Omega_b/\Omega_0\sim0.15h_{50}^{-3/2}/4=0.04h_{50}^{-1}$.
This ratio is marginally consistent with the prediction of cosmic
nucleosynthesis and therefore, survives a ralative large
$\Omega_0$ Universe.
The more examples of arc-like images of background
galaxies associated with clusters of galaxies, expecially at the
outskirts ($\sim1$ Mpc),  will be needed to further confirm this result.\\

After this letter was completed, two groups announced the similar
analysis based on one EMSS selected cluster (MS1224) (Fahlman et al. 1994)
and three Abell clusters (A2218, A1689 and A2163) (Barul \&
Miralda-Escud\'e 1994). They have reached essentially the same conclusions.\\

\bigskip
\bigskip
\bigskip
\bigskip

%----------------------------------------------

{\noindent}{\bf ACKNOWLEDGMENTS}\\
I thank  Francois Hammer for communicating the result of arc
survey in the EMSS selected clusters prior to publication, and
Gray Mamon for helpful discussion.
This work was partially supported by China National Science Foundation.\\

\begin{large}
\begin{center}
{\bf References}\\
\end{center}
\end{large}

\bigskip

{\noindent}Abramopoulos, F., Ku, W.H.-M., 1983, ApJ, 271, 446\\
{\noindent}Arnaud, M. et al., 1992, ApJ, 390, 345\\
{\noindent}Babul, A., Miralda-Escud\'e, J., 1994, preprint\\
{\noindent}Briel, U.G., Henry, J.P., Bohringer, H., 1992, A\&A, 259, L31\\
{\noindent}Carswell, R.F., Rauch, M., Weymann, R.J., Cooke, A.J.,
           Webb, J.K., 1994, MNRAS,\\
	   \null\qquad\quad 268, L1\\
{\noindent}Cavaliere, A., Fusco-Femiano, R., 1876, A\&A, 49, 137\\
{\noindent}Ceballos, M.T., Barcons, X., 1994, preprint\\
{\noindent}Cowie, L.L., Henriksen, M., Mushotzky, R., 1987, ApJ, 317, 593\\
{\noindent}Edge, A.C., Stewart, G.C., 1991, MNRAS, 252, 414\\
{\noindent}Edge, A.C., Stewart, G.C., Fabian, A.C., Arnaud, K.A., 1990,
	   MNRAS, 245, 559\\
{\noindent}Eyles, C.J. et al., 1991, ApJ, 376, 23\\
{\noindent}Fahlman, G., Kaiser, N., Squires, G., Woods, D., 1994, preprint\\
{\noindent}Fort, B., 1992, in Toulouse Workshop on Gravitational Lensing,
           eds. Y.Mellier et al.,\\
           \null\qquad\quad  221\\
{\noindent}Fort, B., Prieur, J.L., Mathez, G., Mellier, Y., Soucail, G.,
           1988, A\&A, 200, L17\\
{\noindent}Gioia, I.M., Henry, J.P., Maccacaro, J., Morris, S.L., Stockes,
           J.T., Wolter, A., 1990,\\
           \null\qquad\quad  ApJ, 356, L35\\
{\noindent}Gioia, I.M., Luppino, G.A., 1994, ApJS, in press\\
{\noindent}Hammer, F., 1991, ApJ, 383, 66\\
{\noindent}Hammer, F., Angonin-Willaime, M.C., Le F\`evre O., Wu, X.P.,
	   Gioia, I.M., Luppino, \\
           \null\qquad\quad G.A., 1993, in the 31 Li\`ege International
           Astrophysical Colloquium \\
           \null\qquad\quad on Gravitational Lenses in the Universe,
           eds. J.Surdej et al., 609\\
{\noindent}Henriksen, M.J., Mamon, G. A., 1994, ApJ, 421, L63\\
{\noindent}Henry, J.P., Briel, U.G., Nulsen, P.E.J., 1993, A\&A, 271, 413\\
{\noindent}Henry, J.P., Gioia, I.M., Maccacaro, T., Morris S.L.,
           Stocke, J.L., Wolter, A.,\\
           \null\qquad\quad  1992, ApJ, 408\\
{\noindent}Henry, J.P., Soltan, A., Briel, U., Gunn, J.E., 1982, ApJ, 262, 1\\
{\noindent}Hughes, J.P., 1989, ApJ, 337, 21\\
{\noindent}Jones, C., Forman, W., 1984, ApJ, 276, 38\\
{\noindent}Le F\`evre O., Hammer, F., Angonin, M.C., Gioia, I.M.,
           Luppino, G.A., 1994,\\
           \null\qquad\quad  ApJ, 422, L5\\
%{\noindent}Loewenstein, M., 1994, ApJ, in press\\
{\noindent}Lubin, L.M., Bahcall N.A., 1993, ApJ, 415, L17\\
{\noindent}Miyaji, T. et al., 1993, ApJ, 419, 66\\
{\noindent}Mulchaey, J.S., Davis, D.S., Mushotzky, R.F., Burstein, D.,
           1993, ApJ, 404, L9\\
{\noindent}Ponman, T.J., Bertram, D., 1993, Nature, 363, 51\\
{\noindent}Soucail, G., Mellier, Y., 1993, in the 31 Li\`ege International
           Astrophysical Colloquium\\
           \null\qquad\quad  on Gravitational Lenses in the Universe,
           eds. J.Surdej et al., 595\\
{\noindent}Songalia, A., Cowie, L.L., Hogan, C.J., Rugers, M.,
           1994, Nature, 368, 599\\
{\noindent}White, S.D.M., Navarro, J.F., Evrard, A.E., Frenk, C.S., 1993,
	   Nature, 366, 429\\
{\noindent}Wu, X.P., Hammer, F., 1993, MNRAS, 262, 187\\
{\noindent}Wu, X.P., 1993, A\&A, 270, L1\\

\end{large}

\newpage

\begin{small}
\begin{tabular}{|c|c|c|c|c|c|c|c|c|c|c|}
\multicolumn{11}{c}
{\Large\bf Table 1 ~~Four arc-cluster systems and their masses}\\
\multicolumn{11}{c}{ }\\
\multicolumn{11}{c}{ }\\
\multicolumn{11}{c}{ }\\
\hline
\multicolumn{3}{|c|}{ } & \multicolumn{5}{|c|}{ } & \multicolumn{3}{|c|}{ }\\
\multicolumn{3}{|c}{\large cluster} & \multicolumn{5}{|c|}{\large arc} &
\multicolumn{3}{c|}{\large mass}\\
\multicolumn{3}{|c|}{ } & \multicolumn{5}{|c|}{ } & \multicolumn{3}{|c|}{ }\\
\hline
          &       &     &    &   &    &          &      &      &      &     \\
name & $z_d$ & $L_{x,44}$ &
$\theta(")$ & $b$ (Mpc) & $L(")$ & $L/W$ & $z_s$ &
 $m_v(b)\;(M_{\odot}) $
 & $m_g(b) \;(M_{\odot})$  & $m_g(b)/m_v(b)$\\
          &       &     &    &   &    &          &      &      &      &     \\
\hline
          &       &     &    &   &    &          &      &      &      &     \\
Abell 370 & 0.374 & 9.7 & 56 & 0.35  & 9 & 18 & 1.3(?)
& $2.26\cdot10^{14}$ & $8.20\cdot10^{14} $ & 3.63\\
          &       &     &    &   &    &          &      &      &      &     \\
\hline
          &       &     &    &   &    &          &      &      &      &     \\
          &       &     &    &   &    &  & 0.6
&  $1.36\cdot10^{14} $ & $6.91\cdot10^{14} $ & 5.08\\
MS1006.0+1202&0.221 & 4.819 & 62 & 0.28 &  4.9&7.0           &      &      &
  &     \\
          &       &     &    &   &    & &  2.0
& $1.36\cdot10^{14} $ & $4.87\cdot10^{14} $ & 3.58\\
          &       &     &    &   &    &          &      &      &      &     \\
\hline
          &       &     &    &   &    &          &      &      &      &     \\
          &       &     &    &   &    & & 0.6
& $1.33\cdot10^{14} $ & $7.47\cdot10^{14} $ & 5.60\\
MS1008.1-1224&0.301& 4.493 & 51 & 0.28 &  4.0&6.5           &      &      &
 &     \\
          &       &     &    &   &    & & 2.0
& $1.33\cdot10^{14} $ & $4.34\cdot10^{14} $ & 3.25\\
          &       &     &    &   &    &          &      &      &      &     \\
\hline
          &       &     &    &   &    &          &      &      &      &     \\
          &       &     &    &   &    &  & 0.6
& $1.63\cdot10^{14} $ & $9.95\cdot10^{14} $ & 6.12\\
MS1910.5+6737&0.246 & 4.386 & 67 & 0.33 &  6.1&10.5          &      &      &
  &     \\
          &       &     &    &   &    &  & 2.0
&  $1.63\cdot10^{14} $ & $6.64\cdot10^{14} $ & 4.08\\
          &       &     &    &   &    &          &      &      &      &     \\
\hline
\end{tabular}
\end{small}
\end{document}